# On the number of independent adiabatic invariants for gyrating particles

O. Ågren, V.E. Moiseenko[1] and A. Gustafsson

Uppsala University, Ångström laboratory, division of electricity and lightning research, Box 534, SE-751 21 Uppsala, Sweden

[1]Institute of Plasma Physics, National Science Center "Kharkov Institute of Physics" and Technology, Akademichna st. 1, 61108 Kharkiv, Ukraine

email: Olov.Agren@uth.uu.se, Phone: +46-18-471 5816 Fax: +46-18-471 3000

email: moiseenk@ipp.kharkov.ua

**Abstract:** PACS numbers: 52.50.Lp, 52.55.Jd, 52.55.Ez

It is pointed out that the three established adiabatic invariants are separating invariants in the sense of Liouville. It is widely claimed that no more than three adiabatic invariants can exist for the motion of a point charge. However, additional independent (not separating) adiabatic invariants do exist. For a force free motion, the components of angular momentum provide two additional constants of motion. This result can be generalized to the Hamilton Jacobi equation. The number of independent constants of motion is reduced if there is a global symmetry. For a gyrating particle, neglecting a gyro helix type of invariant, four "useful" invariants could exist. A radial drift invariant, corresponding to the average of the radial coordinate of the particle, is a constant of motion for a confined gyrating particle. For the special case of a screw pinch where each gyro center moves on a magnetic flux surface without mirror trapping, the radial drift invariant is the radial coordinate $\bar{r}$ of the gyro center. For a screw pinch, the set $(\varepsilon, v, \bar{r})$ of constants of motion is convenient to model the equilibrium. Local Maxwellian distribution functions expressed in this set of invariants are demonstrated to provide MHD-type of equilibria, for which it is straightforward to model the radial profiles of the particle and field components.

# I  Introduction

Constants of motion are of basic importance for plasma physics. Constants of motion can be useful to get an overall picture of the motion, without explicit solutions for the orbits. Jean's theorem also states that an arbitrary function

$$f = f(I_1, I_2, \ldots, I_n)$$

of the motional invariants $I_i(\mathbf{x},\mathbf{v},t)$ is a solution of the Vlasov equation. The motional invariants $I_i(\mathbf{x},\mathbf{v},t)$ are constant along the particle orbits, and may in general depend on time as well as on the phase space variables $(\mathbf{x},\mathbf{v})$. Of particular interest are constants of motion $I_i(\mathbf{x},\mathbf{v})$, which are invariants that do not depend explicitly on time, since stationary equilibrium distribution functions are functions of such invariants. The number of independent constants of motion is thus of significant conceptual importance for basic plasma physics.

A widespread opinion is that at most three independent stationary constants of motion can be found for a point charge. A reason for this belief is that the theorem on integrable systems by Liouville implies that a point charge orbit has *at most three separating* invariants. Six first order differential equations describe the motion. Each separating invariant reduces the Hamiltonian system of equations by two. If two separating invariants other then the energy are found (in a system with two cyclic coordinates, the conserved canonical momenta are two separating invariants), the remaining system is a set of two first order differential equations. In a stationary field, this final set of two first order differential equations have the energy of the particle as the remaining third separating invariant, and the orbit can be determined by quadratures. There are thus at most three separating invariants for the motion of a point charge in a stationary field [1].

Despite that it is impossible for more than three separating invariants to exist, it turns out that *additional independent* constants of motion can be found when the equations of motion are solved [2,3]. It is straightforward to demonstrate this procedure for force free motion, and even the Hamilton Jacobi equation can be treated with the same approach. An invariant is independent of the other constants of motion if it span up a different portion of phase space. Separating invariants are independent invariants in involution, i.e. they satisfy the Poisson bracket relations $\{I_i, I_j\}_{\mathbf{x},\mathbf{p}} = 0$. All independent invariants, also the nonseparating, are required for the complete solution of the Vlasov equation.

The separating invariants are sufficient to integrate the orbits, but, contrary to what is sometimes claimed, this does not imply that the orbit is a function of the separating invariants only. Since the separating invariants result in first order differential equations for the orbit, the orbit is a function of the separating invariants *and a curve parameter*. As pointed out in Ref. 2, by eliminating the curve parameter from the orbit by a function dependent on the phase space variables $(\mathbf{x}, \mathbf{v})$, a new constant of motion is found.

The initial position and velocity provide the six time dependent invariants $x_i(0) = f_i(\mathbf{x}, \mathbf{v}, t)$ and $v_i(0) = g_i(\mathbf{x}, \mathbf{v}, t)$, and all constants of motion should in principle be functions of these invariants which reduce to time independent invariants. G. Schmidth has suggested [4], if only a single phase variable is needed to specify the orbit, that as much as five stationary invariants may be found, see p. 70 in Ref. 3. This works locally near the initial position (an infinitesimal canonical transformation can generate five constants of motion), but the procedure can become problematic globally. We will demonstrate by some simple examples that five constants of motion can be identified in certain cases, but we will also point out that for the gyrating motion in a constant magnetic field, the fifth invariant is a peculiar "gyro helix invariant" which depends on the gyro angle. It does not seem possible, from a process based on an asymptotic series expansion with the gyro radius as a small parameter, to extend this gyro helix invariant to an adiabatic invariant in more general fields with slow gradients. Neglecting the possibility for a fifth invariant for this reason, a pragmatic view is that a system of interest for confining fusion plasma would at most have four independent constants of motion.

The established adiabatic theory [5,6,7], provides the three adiabatic invariants $(\varepsilon, \mu, \tilde{J})$, i.e. the energy, the magnetic moment and a longitudinal invariant, which are separating invariants in the sense of Liouville (the separating adiabatic invariants are determined by an asymptotic expansion) [5,6]. Similar results have been derived by Hamiltonian methods with repeated canonical transformations, initiated with the work by Gardner [8] and developed further in [9-12], compare also [13-15] where non-hamiltonian treatments have been used. At least for systems with proper (nested) fluxsurfaces, the existence of an additional constant of motion can be imagined from the *global* condition that a confined particle cannot escape from the confining magnetic field region, and this implies that the average of the radial Clebsch coordinate $I_r \approx \overline{r_0}(\mathbf{x}, \mathbf{v})$ of the particle along the guiding center orbit has to be constant [3]. The radial invariant $I_r$ is a stationary invariant for Vlasov equilibria [2,3]. To lowest order in the radial displacements of the guiding center, the invariant $I_r$ corresponds to a gyro center motion on a magnetic surface [2,3]. In screw pinches, each gyro center is restricted to move on a single magnetic surface, see Fig. 1. In axisymmetric tori, there is a small correction due to finite banana widths and small radial gyro center excursions [2] [this correction can be calculated from the projected gyro center orbit, which forms a closed curve in the $(r, z)$ plane both for trapped and passing particles [2], see Fig. 2]. The constancy of the mean value of the radial radial coordinate can be applied to magnetic mirrors [16], compare also [14], as well as axisymmetric tori [2,3]. We expect the radial invariant to be independent of $(\varepsilon, \mu, \tilde{J})$ for mirror trapped particles. For screw pinches, where no mirror trapping occurs, it cannot be directly seen whether or not the radial invariant is a fourth independent invariant, and this question will be analyzed in detail.

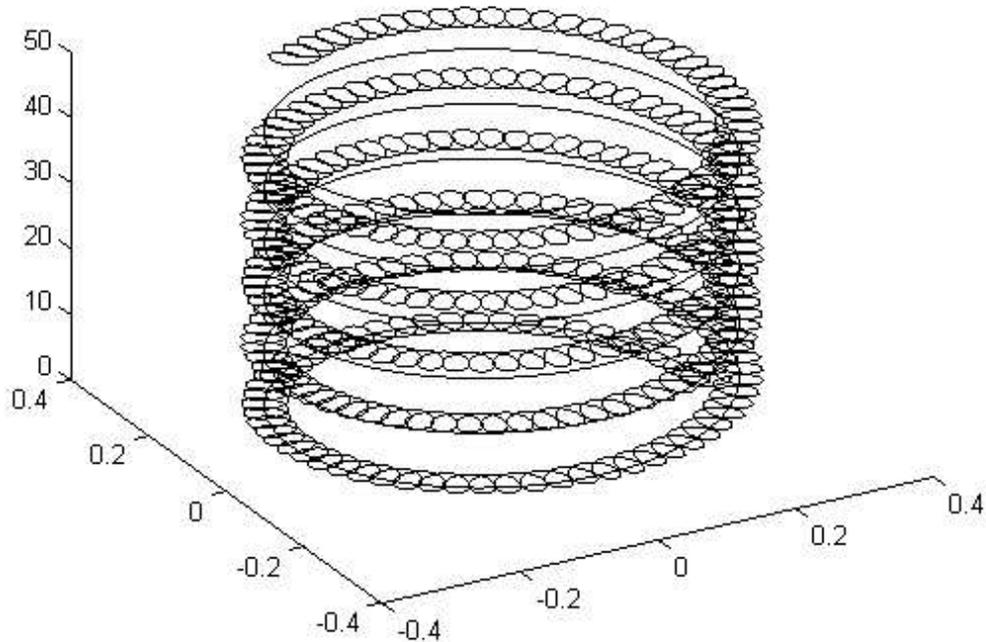

**Fig.1** Particle orbit in a screw pinch. Flux surfaces are cylinders $r = constant$, and each guiding center is restricted to move on a flux surface. A small perpendicular drift gives rice to a slow departure between the magnetic field lines and the guiding center orbit.

Present textbooks in plasma physics claim that at most three adiabatic invariants exist. On the other hand, gyro kinetic equations based on a slow evolution of the gyro center Clebsch coordinates derive in principle the radial drift invariant [10,11,15]. The neoclassical transport theory starts with an assumption that to leading order the gyro center moves on a magnetic surface, and banana orbit widths effects are calculated as a perturbation [17] (the banana widths can still give a substantial contribution to the transport). This approach, and its relation to the collisionless state, is consistent with a radial drift invariant $I_r = \bar{r}_0 + I_r^{(1)}$, which to leading order is equal to the guiding center radial coordinate $\bar{r}_0$ [2]. Although the existence of a radial invariant seems to be nowhere explicitly stated (it is hidden in the algebraic calculations, but follows from the property that a stationary distribution function is in itself a constant of motion), it should be pointed out that gyro kinetic and neoclassical transport theories such as those in Refs. [9-11,17] are in agreement with the existence of the radial drift invariant. In these senses, ideas consistent with a constant radial invariant have been put forward previously. However, we believe it has a general interest for basic plasma physics to explicitly state the existence of a radial invariant, and to analyze its dependence or independence of the three established adiabatic invariants.

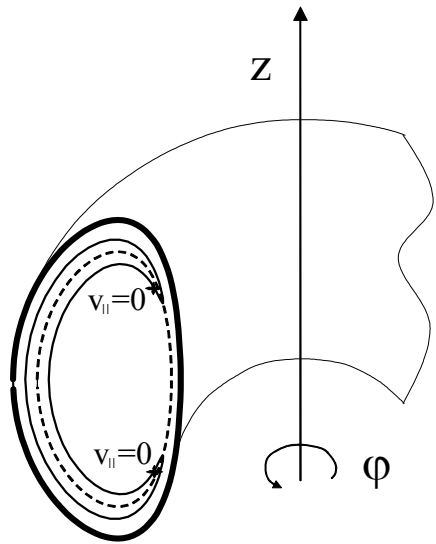

**Fig.2** Gyro center orbits projected on the $(r,z)$ plane of an axisymmetric tori form closed curves for passing as well as trapped particles. Passing particles encircle the magnetic axis with small displacements (typically of the order of a gyro radius) off the initial magnetic flux surface. The crosses in the figure mark schematically positions where the parallel velocity is zero (only for barely trapped particles are these points coinciding with the projected banana tips [2]). The mean value of the radial flux coordinate is an adiabatic invariant which changes continuously as the orbit parameters are changed from a trapped to a passing state.

The existence of a radial invariant has a range of implications. The dependence of the distribution function on the radial invariant determines the radial profiles of the density and temperature and this gives a diamagnetic drift in a direction perpendicular to the magnetic field [11,2,3]. In tokamaks, with the standard set $(\varepsilon, \mu, p_\varphi)$ of three invariants, it is problematical to model the poloidal current [2,3], but the poloidal current can directly be determined by using the radial invariant in the distribution function [11,2]. It is also possible to establish a bridge between Vlasov equilibria and ideal MHD with the use of the radial invariant [2] in a nearly Maxwellian tokamak distribution [2]. The same distribution function also provides a connection between the collsionsless state and the starting point for neoclassical theory, where the guiding center distribution to lowest order is assumed to be a local Maxwellian with constant values for the density and temperature at magnetic flux surfaces [2,17,18].

The main purpose of this paper is to add some clarity to the number of independent invariants. The corresponding question for the three body gravitational motion has been the subject of extensive studies, and in the classical book by Whittaker, see chapter 14 in Ref. [19], it is stated that exactly 10 constants of motions expressed by algebraic functions of the phase space coordinates do exist, however, additional invariants expressed by more general functions could exist. In paragraph 148 in Whittaker, which is restricted to systems with a complete set of separating invariants (invariants in involution), the results correspond to 6 timedependent invariants for a point charge, which could be associated with the initial position and initial velocity of the particle, but there is no indication how this relates to a derivation of the complete set of stationary invariants. For the single particle motion under study in the present paper, it will be shown that the number of first order differential equations for the characteristics is related to this question, and it will be shown how global symmetries are important. In particular, it is addressed how the established adiabatic invariants [5,6,7] relate to the question on the number of independent adiabatic invariants. Explicit results are presented for force free motion, the Hamilton Jacobi equation, the motion in a constant magnet field and the motion in general adiabatic screw pinch equilibria.

## II  Basic mathematical properties of the Vlasov equation

The collisionless Boltzmann equation is a linear first order differential equation in the 6-dimensional phase space variables and the time,

$$(\frac{\partial}{\partial t} + \frac{d\mathbf{x}}{dt} \cdot \frac{\partial}{\partial \mathbf{x}} + \frac{d\mathbf{v}}{dt} \cdot \frac{\partial}{\partial \mathbf{v}}) f(\mathbf{x}, \mathbf{v}, t) = 0 \qquad (1a)$$

where $d\mathbf{x}/dt = \mathbf{v}$ is the velocity and $Md\mathbf{v}/dt = \mathbf{F}$ is the force acting on the particles. In a non dense gas, $d\mathbf{v}/dt = 0$ if collisions are neglected, and the particles move along straight lines. The Vlasov equation corresponds to the case where the acceleration is determined by the Lorentz force, i.e. $Md\mathbf{v}/dt = q(\mathbf{E} + \mathbf{v} \times \mathbf{B})$.

The constants of motion satisfy the stationary collisionless Bolzmann equation

$$\sum_{i=1}^{3} (\mathrm{v}_i \frac{\partial f}{\partial x_i} + \frac{F_i}{M} \frac{\partial f}{\partial \mathrm{v}_i}) = 0 \qquad (1b)$$

where $\partial \mathbf{F} / \partial t = 0$. This is a first order differential equation, and mathematical theory tells us that its solutions are fully determined by the characteristics which could be derived from the set of five ordinary differential equations

$$\frac{dx}{v_x} = \frac{dy}{v_y} = \frac{dz}{v_z} = \frac{M dv_x}{F_x} = \frac{M dv_y}{F_y} = \frac{M dv_z}{F_z} \tag{1c}$$

A solution trajectory of Eq. (1c) can be parameterized by a curve parameter $\tau$, and the equations for the characteristics of the stationary Vlasov equation can be written $d\mathbf{x}(\tau)/d\tau = \mathbf{v}(\tau)$ and $M d\mathbf{v}(\tau)/d\tau = \mathbf{F}[\mathbf{x}(\tau), \mathbf{v}(\tau)]$. This is the well known result that the equations of the characteristics of the collsionless Boltzmann equation are determined by the particle orbits. The form of the five equations in (1c) determine directly (since the time is eliminated from the equations) five time independent quantities that depend on the phase space variables $(\mathbf{x}, \mathbf{v})$.

A suitable set of phase variables should be selected for Eq. (1b), and we restrict the analysis to a finite domain with a finite variation of the phase space variables. By restricting the analysis to that kind of finite domain, certain intricate mathematical questions such as multi-valued functions can be avoided. Mathematical theory tells, see the pages 3-62 in Ref. [20], that if the derivatives $\partial F_i / \partial x_j$ are continuous in the domain, five independent solutions of the system (1c) of five equations exist in the domain and could be written as

$$I_k(\mathbf{x}, \mathbf{v}) = const \tag{2a}$$

where $k = 1,..,5$. Each such invariant satisfies the kinetic equation (1b). The functional independence of these constants of motion is proved by the fact that the functional matrix $\partial (I_1, ... I_5) / \partial (\mathbf{x}, \mathbf{v})$ has a rank 5. Additionally if $f_1$ is also a solution of the kinetic equation, then

$$\left|\frac{\partial(I_1,...I_5,f_1)}{\partial(\mathbf{x},\mathbf{v})}\right| = 0,$$

i.e. the Jacobian nullifies proving that a functional dependence exists and no more independent solution is possible. In this way 5 independent constants of motion $I_k(\mathbf{x},\mathbf{v})$ could be found for the kinetic equation. From this follows that any stationary solution of the kinetic equation for point masses is a function of the five independent constants of motion, i.e.

$$f = f(I_1, I_2, I_3, I_4, I_5).  \tag{2b}$$

The original version of Jeans theorem is that the distribution function is a function of an unspecified number of invariants, but here the theorem is strengthened to the more specific result that the general solution is a function of five independent constants of motion.

For systems with a global symmetry, i.e. one or two coordinates are cyclic for all particles, a special treatment of the characteristic equations is required. For an azimuthally and longitudinally uniform screw pinch the kinetic equation reads in cylindrical coordinates,

$$v_r \frac{\partial f}{\partial r} + \left(\frac{F_r}{M} + \frac{v_\theta^2}{r}\right)\frac{\partial f}{\partial v_r} + \left(\frac{F_\theta}{M} - \frac{v_r v_\theta}{r}\right)\frac{\partial f}{\partial v_\theta} + \frac{F_z}{M}\frac{\partial f}{\partial v_z} = 0 \tag{3a}$$

where the global symmetries $\partial f/\partial \theta = \partial f/\partial z = 0$ are imposed. The characteristic equations are then

$$\frac{dr}{v_r} = \frac{Mdv_r}{q(E_r + v_\theta B_z - v_z B_\theta) + \frac{Mv_\theta^2}{r}} = \frac{-Mdv_\theta}{qv_r B_z + \frac{Mv_r v_\theta}{r}} = \frac{Mdv_z}{qv_r B_\theta} \tag{3b}$$

This system of three equations have the three integrals $(\varepsilon, p_\theta, p_z)$, i.e. the energy and two canonical momenta associated with the two cyclic coordinates. Since the system of equations (3b) is reduced to only three equations by the imposed symmetries, the theory in Ref. [20] shows that these three integrals are the complete system of constants of motion for Eq. (3a). An analogous conclusion holds for slab geometry with only one non-ignorable coordinate.

In axisymmetric tori, where only the toroidal angle is a cyclic coordinate, the characteristic equations is a system of four first order ordinary differential equations, and there could in principle exist 4 independent constants of motion. (In passing, it can be noticed that the global symmetry can be destroyed by the approximating transformation procedure introduced by Gardner[8]; however Weitzner[9] have developed the Hamiltonian method further to avoid this problem). It has been shown in Ref. 2 that a radial invariant can be determined from the projected gyro center orbit, which is a closed orbit in the (r,z) plane for both trapped and passing particles. For the special case of a screw pinch, Eqs. (3a,b) show that the radial invariant is a dependent invariant, i.e. $I_r = I_r(\varepsilon, p_\theta, p_z)$. With a finite major radius, banana trapped particles appear, which has no correspondence in a screw pinch, and we expect the radial invariant to be independent of the three standard constants of motion $(\varepsilon, \mu, p_\theta)$ of an axisymmetric tori.

For a force free-motion, the characteristic equations are

$$\frac{dx}{v_x} = \frac{dy}{v_y} = \frac{dz}{v_z} = \frac{dv_x}{0} = \frac{dv_y}{0} = \frac{dv_z}{0}$$

In this case, the three velocity components and two components of the angular momentum provide a set of 5 independent integrals, as will be shown in Section III.

Let us return to the screw pinch and investigate this case in more detail. The Hamiltonian version of the Vlasov equation,

$$\frac{\partial f(\mathbf{q},\mathbf{p},t)}{\partial t} - \{H, f\}_{\mathbf{q},\mathbf{p}} = \frac{\partial f}{\partial t} + \frac{\partial H}{\partial \mathbf{p}} \cdot \frac{\partial f}{\partial \mathbf{q}} - \frac{\partial H}{\partial \mathbf{q}} \cdot \frac{\partial f}{\partial \mathbf{p}} = 0$$

is invariant under canonical transformations. The Hamiltonian structure shows that if a separating invariant is found, the system of first order differential equations can be reduced in the same manner as for the single particle motion. For a stationary screw pinch, we can use cylindrical coordinates $(r,\theta,z)$ and that $(p_\theta, p_z, \varepsilon)$ are separating invariants if the Hamiltonian does not depend on $\theta$, $z$ and $t$. The bounce time for the radial motion is

$$\tau(r, p_\theta, p_z, \varepsilon) = \int^r \frac{dr'}{v_r(r')} = \sigma_r \sqrt{\frac{M}{2}} \int^r \frac{dr'}{\sqrt{\varepsilon - U(r', p_\theta, p_z)}}$$

where $\sigma_r$ is the sign of the radial velocity and $U(r, p_\theta, p_z)$ is the effective potential for the radial motion. The generating function

$$G(r, \theta, z, \varepsilon, p_\theta, p_z) = \theta p_\theta + z p_z + \sigma_r \sqrt{2M} \int^r dr' \sqrt{\varepsilon - U(r', p_\theta, p_z)}$$

of old canonical coordinates and the new momenta $(p_\theta, p_z, \varepsilon)$ provides a canonical transformation to the new coordinates $(Q_1, Q_2, \tau)$. We have $\tau \equiv \partial G / \partial \varepsilon$ and using $\dot\theta = \partial U / \partial p_\theta$ and $\dot z = \partial U / \partial p_z$ the other two new canonical coordinates become

$$Q_1 \equiv \frac{\partial G}{\partial p_\theta} = \theta - \int_0^{\tau(r)} d\tau' \frac{\partial U[r(\tau'), p_\theta, p_z]}{\partial p_\theta} = \theta_{in}$$

$$Q_2 \equiv \frac{\partial G}{\partial p_z} = z - \int_0^{\tau(r)} d\tau' \frac{\partial U[r(\tau'), p_\theta, p_z]}{\partial p_z} = z_{in}$$

This is the initial values of $\theta$ and $z$. Obviously, $Q_1$ and $Q_2$ are constant and these canonical variables depend on $\theta$ and $z$. The stationary Vlasov equation reads in the new canonical variables

$$(\frac{\partial f}{\partial \tau})_{Q_1, Q_2, p_\theta, p_z, \varepsilon} = 0$$

Formally, for an effective potential $U(r, p_\theta, p_z)$ the general solution of the Vlasov equation $F(p_\theta, p_z, \varepsilon, Q_1, Q_2)$ is a function of five constants of motion, which is in agreement with the result in Eq. (2). The dependence on the canonical coordinates $(Q_1, Q_2)$ demonstrates a solution that is valid in a larger phase space domain than the restricted phase space where two cyclic coordinates are filtered out from the analysis. Although it is possible to find more general stationary solutions of the Vlasov equation than functions solely of the separating invariants, the stationary solution of the Vlasov equation and the Maxwell equations should be consistent with an effective potential $U(r, p_\theta, p_z)$ and fields satisfying the global symmetries, which in the transformed coordinates implies that the field quantities are independent of the canonical coordinates $(Q_1, Q_2)$. If we impose the condition that also the distribution function (not only the Hamiltonian) should be independent on the cyclic coordinates $\theta$ and $z$, the distribution function in a stationary screw pinch is of the form $F(\varepsilon, p_\theta, p_z)$.

### III Constants of motion in force free motion

There exists 5 independent stationary constants of motion $I_i = I_i(\mathbf{x}, \mathbf{v})$ in this case. The three velocity components are separating invariants, the particle moves along a straight line where the initial position $(x_{in}, y_{in}, z_{in})$ determines the three timedependent invariants,

$$x_{in}(\mathbf{x}, \mathbf{v}, t) = x - \mathrm{v}_x t$$

$$y_{in}(\mathbf{x}, \mathbf{v}, t) = y - \mathrm{v}_y t$$

$$z_{in}(\mathbf{x}, \mathbf{v}, t) = z - \mathrm{v}_z t$$

Invariants dependent only on the $(\mathbf{x}, \mathbf{v})$ variables can be constructed by eliminating the explicit time dependence from the invariants above. This leads to three components of the angular momentum of the particle,

$$I_4(\mathbf{x}, \mathbf{v}) = x\mathrm{v}_y - y\mathrm{v}_x \tag{4a}$$

$$I_5(\mathbf{x}, \mathbf{v}) = x\mathrm{v}_z - z\mathrm{v}_x \tag{4b}$$

and $I_6(\mathbf{x}, \mathbf{v}) = y\mathrm{v}_z - z\mathrm{v}_y$. This sixth constant of motion is a function of the other five in view of the relation $\mathrm{v}_z I_4 - \mathrm{v}_y I_5 + \mathrm{v}_x I_6 = 0$. For a force free motion, there exists a set of *five* independent constants of motion $I_i = I_i(\mathbf{x}, \mathbf{v})$,

$$(\mathrm{v}_x, \mathrm{v}_y, \mathrm{v}_z, I_4, I_5) \tag{4c}$$

From the six invariants corresponding to the initial position and the initial velocity, five stationary constants of motion can be constructed, while one variable (curve parameter) is required to parameterize the motion from the initial postion. This result is in agreement with the suggestion by G. Schmidt [4].

Any distribution function $F(\mathbf{v})$ that depends only on the velocity components give rice to a constant density. The angular momenta provide spatial localization of the particle distribution. This can be illustrated by the distribution function

$$F_i = n_0 \frac{\exp[-(\mathrm{v}^2 + J_4^2)/\mathrm{v}_{th,i}^2]}{(\pi \mathrm{v}_{th,i}^2)^{3/2}}, \tag{5a}$$

where $J_4 = (x\mathrm{v}_y - y\mathrm{v}_x)/a$, $n_0$, $\mathrm{v}_{th}$ and $a$ are parameters and the index $i$ stands for the ion and electron components. For a finite scale length $a$, a density peaked at $x = y = 0$ is obtained,

$$n(x, y) = \frac{n_0}{\sqrt{1 + \frac{x^2 + y^2}{a^2}}} \tag{5b}$$

for both ions and electrons. A similar calculation shows that there is no current, i.e. $\mathbf{j} = 0$. With this distribution function for both ions and electrons, a quasi-neutral calculation shows that the plasma give rice to neither an electric field nor a magnetic field. The agreement with Maxwell equations are therefore exact (apart from special relativity effects, which are neglected), thus proving the possibility to obtain a consistent description with a distribution function dependent on more invariants than the three separating invariants. This consistency is obvious if the distribution function is aimed to describe the force free motion of atoms or molecules in an ordinary gas.

**IV  Constants of motion in the Hamilton Jacobi equation**

The five invariants may seem like an exceptional case. However, a large number of studies of gyrating particle motion have been devoted to find a canonical transformation (by some asymptotic expansion) to action angle variables, where the transformed Hamiltonian $H = H(J_1, J_2, J_3)$ is a function of the action variables only [1]. The idea behind the action angle formalism is to transform the resulting orbit equations into the straight lines $Q_i(t) = Q_i(0) + \omega_i(\mathbf{J})t$ for the transformed canonical coordinates, where the frequencies $\omega_i(\mathbf{J}) = \partial H / \partial J_i$ are constant. In case $\omega_i(\mathbf{J}) \neq 0$, two independent constants of motion, in addition to the separating invariants $(J_1, J_2, J_3)$ are the "generalized angular momenta"

$$I_4(\mathbf{x}, \mathbf{v}) = Q_1 \omega_2(\mathbf{J}) - Q_2 \omega_1(\mathbf{J}) \qquad (6a)$$

$$I_5(\mathbf{x}, \mathbf{v}) = Q_1 \omega_3(\mathbf{J}) - Q_3 \omega_1(\mathbf{J}) \qquad (6b)$$

while a sixth dependent invariant satisfies $0 = \omega_3 I_4 - \omega_2 I_5 + \omega_1 I_6$. Thus five independent stationary invariants can be instructed from the Hamilton Jacobi equations, exactly in the same manner as for the constant velocity case. However, the derivation assumes $\omega_i(\mathbf{J}) \neq 0$, and this no degeneracy condition need not be satisfied, in particular if there are two global cyclic coordinates. For the screw pinch, we have seen that $\dot{Q}_1 = \omega_1(\mathbf{J}) = 0$ and $\dot{Q}_2 = \omega_2(\mathbf{J}) = 0$, and for a screw pinch there are only three constants of motion which are independent of the cyclic coordinates $\theta$ and $z$.

In most plasma physics text books, it is claimed that there exists at most three adiabatic invariants. This claim is not correct. A deeper look into the work by Kruskal [5], who developed a systematic scheme to construct adiabatic invariants, reveals that the adiabatic invariants considered are *separating* invariants [5] for the motion of an individual particle. It is true that no more than three separating (adiabatic) invariants could be found for the motion of a point charge [1], but a fourth (or even fifth for certain systems) adiabatic invariant may exist, in addition to the separating invariants [2,3].

It is of interest to describe the method of Kruskal in some more detail. (It could be mentioned that Gardner seems to have been the first to publish an approximating scheme to reduce the order of the Hamiltonian system of equations[8,9], and the Kruskal method have some similarities with that approach. Gardner also outlined a means to determine an adiabatic longitudinal invariant[8,9], but the first explicit demonstration of the conditions for its constancy in average was carried out by Northrop and Teller[6,7]). The Kruskal method has widely been interpreted as a construction of all possible adiabatic invariants [5,6,7]. A closer look reveals that only the adiabatic separating invariants are derived by the method. For a gyrating point charge in a stationary field, the degree of freedom is $N = 3$. Kruskal presents a scheme to construct *separating* invariants (in an asymptotic series sense, assuming adiabatic particle motion), whereby the order of the Hamiltonian system is reduced when the separating adiabatic invariant is found [5,6]. The first adiabatic separating invariant is the Poincare invariant, which is an integral of the form [5,6]

$$I_1 = \int \mathbf{p} d\mathbf{q} = \sum_{i=1}^{N=3} \int p_i dq_i$$

A canonical transformation $(\mathbf{q},\mathbf{p}) \to (\mathbf{Q},\mathbf{P})$, where $P_3 = I_1$ and the associated canonical coordinate $Q_3$ is a cyclic coordinate for the particle in an asymptotic series sense, is performed to reduce the order of the problem. The invariant $I_1$ is basically the magnetic moment, while $Q_3$ to leading order is the gyro angle. The canonical transformation reduces the Hamiltonian to the form $H = H(Q_1, Q_2, P_1, P_2, I_3)$, which corresponds to a set of four first order differential equations for $(Q_1, Q_2, P_1, P_2)$. If the motion in these coordinates is nearly periodic, the procedure is repeated, and the second separating adiabatic invariant is the Poincare invariant [5,6],

$$I_2 = \sum_{i=1}^{2} \int P_i dQ_i$$

This invariant is basically the parallel invariant. The second adiabatic invariant can be used in a canonical transformation $(Q_1, Q_2, P_1, P_2) \to (\tilde{Q}_1, \tilde{Q}_2, \tilde{P}_1, \tilde{P}_2)$, where $\tilde{P}_2 = I_2$ and $\tilde{Q}_2$ is a cyclic coordinate for the motion of the considered particle, to reduce the Hamiltonian into the one-dimensional form $H = H(\tilde{Q}_1, \tilde{P}_1, I_2, I_3)$. The remaining problem has the energy as the last separating invariant [5,6].

In passing, it can be noticed that separating coordinates need not be unique. The Kruskal coordinates generate for each particle a set of canonical variables, and relies on adiabaticity but does not require global symmetries. The rapid gyration of each particle is first separated and thereafter the slower oscillatory part of the longitudinal motion. The Kruskal coordinates are applicable for gyrating particles in general adiabatic fields. For the special case of a screw pinch, standard cylindrical coordinates provides an alternative set of separating coordinates.

A critical point is that Kruskal constructs *separating* invariants [5], and the separating invariants can be at most three for a point charge. This does not exclude that a fourth (or even a fifth in some cases) adiabatic invariant can be found [2,3]. As we have seen, Hamilton Jacoby theory may give two additional constants of motion, namely two "generalized angular momenta". For a gyrating particle, one of these should be expected to have a similar phase space dependence as the "gyro helix invariant" for the constant magnetic field case (this will be shown in the next section), and it is even questionable if the asymptotic expansion scheme would be accurate enough to resolve such an invariant. In any case, one other adiabatic invariant in addition to the separating adiabatic invariants could be expected to exist for typical electromagnetic fields [2,3]. In fact, there are a number of examples where it is possible to explicitly identify additional adiabatic invariants, for instance for the force free motion, the Hamilton Jacobi equation, planetary Kepler motion [19], the motion in a constant magnetic field, the motion in magnetic mirrors [16] and trapped particle motion in axisymmetric tori with stationary fields [2].

## V  Constants of motion in a constant magnetic field

The Lorentz force acting on a particle in a constant magnetic field $\mathbf{B} = B_0 \hat{\mathbf{z}}$ is

$$\frac{1}{\Omega_0} \frac{d\mathbf{v}}{dt} = \mathbf{v} \times \hat{\mathbf{z}}$$

where $\Omega_0 = qB_0/M$ is the gyro frequency. We directly obtain the well known four constants of motion $(v_\parallel, v_\perp, \bar{x}, \bar{y})$, i.e. the parallel velocity, the magnitude of the perpendicular velocity and two coordinates for the guiding center position,

$$\bar{x}(\mathbf{x},\mathbf{v}) = x + \frac{v_y}{\Omega_0} \tag{7a}$$

$$\bar{y}(\mathbf{x},\mathbf{v}) = y - \frac{v_x}{\Omega_0} \tag{7b}$$

Bars are used in this manuscript to denote gyro center quantities. The particle orbit is the well known helix

$$z = z_{in} + v\, t$$

$$x = \bar{x} + \frac{v_\perp}{\Omega_0} \cos(\varphi_{g,0} + \Omega_0 t)$$

$$y = \bar{y} - \frac{v_\perp}{\Omega_0} \sin(\varphi_{g,0} + \Omega_0 t)$$

where $\varphi_g = \varphi_{g,0} + \Omega_0 t$ is the gyro angle and the constant $\varphi_{g,0}$ determines the phase of the gyro motion. A fifth independent constant of motion is the "gyrohelix invariant"

$$I_5(\mathbf{x},\mathbf{v}) = v_z \varphi_g - z\Omega_0 \tag{7c}$$

However, this invariant, which depends on the gyro angle, should not be used for typical Vlasov equilibria, since "gyro ripples" on a short length scale would arise if this invariant would be used in a distribution function.

Since $M\Omega_0 \{\bar{x},\bar{y}\}_{\mathbf{x},\mathbf{p}} = -1 \neq 0$ (the invariants are not in involution), we have not determined a set of separating invariants. A vector potential $\mathbf{A}_{cyl} = \hat{\boldsymbol{\theta}} B_0 r / 2$ corresponds to an axisymmetric system, and the energy and the canonical momenta $p_\theta$ and $p_z$ are then separating invariants which are related to the constants of motion $(v, v_\perp, \bar{x}, \bar{y})$ by

$$\varepsilon = \frac{M}{2}(v_\perp^2 + v^2) \tag{8a}$$

$$p_\theta = \frac{M}{2\Omega_0}[(\bar{x}^2 + \bar{y}^2)\Omega_0^2 - v_\perp^2] \tag{8b}$$

$$p_z = Mv \tag{8c}$$

The constancy of $p_\theta$ is related to the specific choice of gauge for the vector potential, and in Eq. (8b) it is implicitly assumed that the constant field region is restricted to a region bounded by an axisymmetric cylinder. If the constant field region would be bounded by a noncircular curve in the *xy* plane, it is typically not possible to find a separating global invariant (that reduces to $p_\theta$ in the axisymmetric case), but it can be noticed that $(v, v_\perp, \bar{x}, \bar{y})$ (as well as the gyro helix invariant) is still valid as a set of constants of motion. Neglecting the particle contributions to the fields, it is straightforward to construct nonaxisymmeric equilibria with various spatial dependences in the *xy* plane by selecting a distribution function $F(v, v_\perp, \bar{x}, \bar{y})$ of four invariants with appropriate dependencies on the spatially localizing invariants $\bar{x}$ and $\bar{y}$.

## VI  Constants of motion in a screw pinch

We will show how the set $(\varepsilon, p_\theta, p_z)$ of separating invariants in a screw pinch can be replaced by the constants of motion $(\varepsilon, v, \bar{r})$. With this latter set, it is more convenient to construct Vlasov equilibria with detailed control of radial profiles for the field components.

In a screw pinch, the guiding center values of the parallel velocity and the radial coordinate are constants of motion, and we will show how these invariants depend on the set of invariants $(\varepsilon, p_\theta, p_z)$ by implicit functions of the form $\bar{r}(p_\theta, p_z)$ and $\bar{v}(p_\theta, p_z)$. For this reason we consider the effective potential for the radial motion

$$\frac{U(r)}{M} = \frac{1}{2r^2}(\frac{p_\theta}{M} - \frac{q\psi}{M})^2 + \frac{1}{2}(\frac{p_z}{M} - \frac{qA_z}{M})^2 + \frac{q\phi}{M}$$

where $Mrv_\theta = p_\theta - q\psi$, $Mv_z = p_z - qA_z$, $rB_z = d\psi/dr$ and $B_\theta = -dA_z/dr$. We will use a harmonic oscillator approximation, where this potential has a minimum at the guiding center radial coordinate

$$\bar{r} = \bar{r}(p_\theta, p_z), \qquad (9a)$$

where $\bar{r} = \bar{r}(p_\theta, p_z)$ is implicitly determined by

$$0 = -\frac{\bar{v}_\theta^2}{\bar{r}} - \Omega_z(\bar{r})\bar{v}_\theta + \Omega_\theta(\bar{r})\bar{v}_z - \frac{q}{M}E_r(\bar{r}) \qquad (9b)$$

where $\Omega_z = qB_z/M$ and $\Omega_\theta = qB_\theta/M$. The first term is typically small if $\bar{r}$ is a few gyro radii or larger (but it is not small for particles encircling the axis). At the minimum of the effective potential, we obtain with $\bar{v} = \hat{\mathbf{B}} \cdot \mathbf{v}$

$$\bar{v}(p_z, p_\theta) = \frac{B_z(\bar{r})}{B(\bar{r})}\frac{p_\theta - q\psi(\bar{r})}{M\bar{r}} + \frac{B_\theta(\bar{r})}{B(\bar{r})}\frac{p_z - qA_z(\bar{r})}{M} \qquad (9c)$$

where $\bar{v}(p_\theta, p_z) = \bar{v}[p_\theta, p_z, \bar{r}(p_\theta, p_z)]$ and $\bar{r} = \bar{r}(p_\theta, p_z)$ is determined by Eq. (9b).

The separating invariants $(\varepsilon, p_\theta, p_z)$ in a screw pinch can thus be replaced by the constants of motion

$$(\varepsilon, \bar{v}, \bar{r}) \qquad (10)$$

where $\bar{r}(p_\theta, p_z)$ and $\bar{v}(p_\theta, p_z)$ are implicit functions determined from the minimum of the effective potential. In passing, we notice that the sign of the parallel velocity,

$$\sigma(\mathbf{x}, \mathbf{v}) = \frac{\bar{v}}{\sqrt{\frac{2}{M}[\varepsilon - q\phi(\bar{r}) - \mu B(\bar{r})]}} \qquad (11)$$

is constant, since no magnetic mirror trapping appears in a screw pinch. Obviously, $\sigma(\mathbf{x}, \mathbf{v})$ is a function of the constants of motion $(\varepsilon, \mu, \bar{v}, \bar{r})$.

A harmonic oscillator approximation can be made around the minimum of the effective potential in a similar way as in Ref. [21]. The frequency of the radial oscillations is determined by

$$\Omega_r^2 = \frac{1}{M}\frac{d^2U}{d\bar{r}^2} = \Omega^2 + \frac{3\bar{v}_\theta^2}{2\bar{r}^2} + (2\frac{\Omega_z}{\bar{r}} - \frac{d\Omega_z}{d\bar{r}})\bar{v}_\theta + \frac{d\Omega_\theta}{d\bar{r}}\bar{v}_z + \frac{q}{M}\frac{d^2\phi}{d\bar{r}^2} \quad (12)$$

where $\Omega^2 = \Omega_z^2 + \Omega_\theta^2$. It is sufficient to use the approximation $\Omega_r = \Omega(\bar{r})$. The energy conservation for the particle becomes in the harmonic oscillator approximation

$$\varepsilon = \frac{Mv_r^2}{2} + U(\bar{r}) + \frac{\Omega_r^2}{2}(r - \bar{r})^2 \quad (13)$$

The solution describes an oscillatory motion around the guiding center coordinate

$$r = \bar{r} + \frac{1}{\Omega_r}\sqrt{\frac{\varepsilon - U(\bar{r})}{M/2}}\cos(\Omega_r t + \varphi_{ro}) = \bar{r} + r_g(\mathbf{x}, \mathbf{v}) \quad (14)$$

where $\varphi_{ro}$ is constant. A check shows that this gives a constant value of

$$\mu = \frac{Mv_g^2}{2B} = \frac{M}{2B}[v_r^2 + (v_\theta - \bar{v}_\theta)^2 + (v_z - \bar{v}_z)^2]$$

where $v_g$ is the gyrating part of the velocity and

$$v_r^2 = \frac{\varepsilon - U(\bar{r})}{M/2} - \Omega_r^2 r_g^2(\mathbf{x}, \mathbf{v})$$

$$v_\theta - \bar{v}_\theta = -(\Omega_z + \frac{\bar{v}_\theta}{\bar{r}})r_g(\mathbf{x}, \mathbf{v}) \approx -\Omega_z r_g(\mathbf{x}, \mathbf{v})$$

$$v_z - \bar{v}_z = \Omega_\theta r_g(\mathbf{x}, \mathbf{v})$$

With $\Omega_r \approx \Omega$, we obtain a constant magnetic moment:

$$\mu = \frac{\varepsilon - U(\bar{r})}{B(\bar{r})} = constant \quad (15)$$

where

$$U(\bar{r}) = \frac{[p_\theta - q\psi(\bar{r})]^2}{2M\bar{r}^2} + \frac{[p_z - qA_z(\bar{r})]^2}{2M} + q\phi(\bar{r})$$

This is an implicit function of the form $U(p_\theta, p_z) = U[p_\theta, p_z, \bar{r}(p_\theta, p_z)]$, and Eq. (15) shows that the magnetic moment is a function of the three separating invariants, i.e.

$$\mu(\varepsilon, p_\theta, p_z) = \frac{\varepsilon - U(p_\theta, p_z)}{B(p_\theta, p_z)}$$

The magnetic moment is therefore a dependent invariant in screw pinches. Similarly, Eq. (11) shows that $\sigma(\mathbf{x}, \mathbf{v}) = \sigma(\varepsilon, p_\theta, p_z)$ is an implicit function of the three separating invariants.

The relation $\mu B(\bar{r}) = \varepsilon - U(\bar{r})$ can be caste into a more transparent form to detect the behavior of the parallel motion. The guiding center velocity $\overline{\mathbf{v}}(\overline{\mathbf{x}}) = \overline{v} \, \hat{\mathbf{B}} + \overline{\mathbf{v}}_\perp(\overline{\mathbf{x}})$ has parallel and perpendicular components. In a screw pinch, the perpendicular drift is of the form $\overline{\mathbf{v}}_\perp(\overline{\mathbf{x}}) = \overline{v}_\perp(\bar{r})\hat{\mathbf{r}} \times \hat{\mathbf{B}}$, and this implies for the components along $\hat{\theta}$ and $\hat{\mathbf{z}}$,

$$\overline{v}_\theta = \frac{B_\theta}{B}\overline{v} - \frac{B_z}{B}\overline{v}_\perp \tag{16a}$$

$$\overline{v}_z = \frac{B_z}{B}\overline{v} + \frac{B_\theta}{B}\overline{v}_\perp \tag{16b}$$

The effective potential at its minimum is $U(\bar{r}) = q\phi(\bar{r}) + M(\overline{v}^2 + \overline{v}_\perp^2)/2$, whereby the energy conservation becomes

$$\varepsilon = M\frac{\overline{v}^2 + \overline{v}_\perp^2}{2} + q\phi(\bar{r}) + \mu B(\bar{r}) \approx M\frac{\overline{v}^2}{2} + q\phi(\bar{r}) + \mu B(\bar{r}) \tag{17}$$

where the perpendicular drift is assumed small in the last step and $q\phi(\bar{r}) + \mu B(\bar{r})$ is the guiding center potential. This shows the well known result that the guiding center parallel velocity is constant in a screw pinch, and this relation was already used in the formula for $\sigma(\mathbf{x}, \mathbf{v})$. The standard guiding center drift formula, obtained by a gyro averaging procedure, for a stationary field is

$$\overline{\mathbf{v}}_\perp(\overline{\mathbf{x}}) \;=\; \frac{\mathbf{E}\times\mathbf{B}}{B^2} + \frac{\mu}{q}\frac{\mathbf{B}\times\nabla B}{B^2} + \frac{M\overline{\mathrm{v}}^2}{qB^4}\mathbf{B}\times[(\mathbf{B}\cdot\nabla)\mathbf{B}] \tag{18}$$

and the corresponding formula for $\overline{\mathbf{v}}_\perp(\overline{r}) = \overline{\mathbf{v}}_\perp(\overline{r},\mu,\overline{\mathrm{v}})$ in a screw pinch becomes

$$\overline{\mathbf{v}}_\perp(r) \;=\; \frac{E_r}{B} - \frac{\mu}{q}\frac{1}{B}\frac{\partial B}{\partial r} + \frac{B_\theta^2}{B^2}\frac{\overline{\mathrm{v}}^2}{r\Omega} \tag{19}$$

As a comparison, we notice that Eqs. (9b) and (16a,b) yield at the minimum of the effective potential

$$\overline{\mathbf{v}}_\perp \;=\; \frac{E_r}{B} + \frac{(B_\theta\overline{\mathrm{v}} - B_z\overline{\mathrm{v}}_\perp)^2}{B^2\overline{r}\Omega} \;\approx\; \frac{E_r}{B} + \frac{B_\theta^2}{B^2}\frac{\overline{\mathrm{v}}^2}{r\Omega}$$

which is in qualitative agreement with Eq. (19) if the $\mathbf{E}\times\mathbf{B}$ drift is the dominating term for the perpendicular drift. More precisely, in a screw pinch the last equation reduces to Eq. (19) when $|B_\theta\overline{\mathrm{v}}|$ $|B_z\overline{\mathrm{v}}_\perp| \approx |E_r|$ and the $\nabla B$ drift is small compared to the curvature drift, which corresponds to $|\mathrm{v}_\perp^2 B_\theta^2 r\partial B/\partial r| \ll \overline{\mathrm{v}}^2 B^3$, or roughly $|B_\theta^2 r\partial B/\partial r| \ll B^3$. This latter condition determines a limitation on the field gradients caused by the finite beta currents. For the special case of a $z$ pinch without an axial field, this inequality reduces to $|r\partial B/\partial r| \ll B$, which is not satisfied near the axis, nor asymptotically.

With a small perpendicular drift, we obtain $\varepsilon - U(\overline{r}) = \mu B(\overline{r}) \approx M\overline{\mathrm{v}}_\perp^2/2$, whereby

$$r \;=\; \overline{r} + \frac{\mathrm{v}_\perp}{\Omega(\overline{r})}\cos\varphi_g \;=\; \overline{r} + r_g(\mathbf{x},\mathbf{v}) \tag{20a}$$

where $\varphi_g = \Omega(\overline{r})t + \varphi_{ro}$ is the gyro angle and $\mathrm{v}_\perp/\Omega(\overline{r})$ is the small gyro radius. The evolution of the coordinates $\theta$ and $z$ along the particle orbit can be determined from a small gyro radius expansion of $Mr^2\dot\theta = p_\theta - q\psi$ and $M\mathrm{v}_z = p_z - qA_z$. This result in

$$r\frac{d\theta}{dt} \;=\; \overline{\mathrm{v}}_\theta - (\Omega_z + \frac{\overline{\mathrm{v}}_\theta}{r})r_g(\mathbf{x},\mathbf{v}) \;\approx\; \overline{\mathrm{v}}_\theta - \Omega_z(\overline{r})r_g(\mathbf{x},\mathbf{v}) \tag{20b}$$

$$\frac{dz}{dt} = \overline{v}_z + \Omega_\theta(\overline{r}) r_g(\mathbf{x}, \mathbf{v}) \qquad (20c)$$

where the first terms on the RHS are the guiding center velocity terms. A result of these equations is

$$v_\parallel \approx \overline{v}_\parallel \qquad (21)$$

The parallel velocity of the particle and the guiding center are thus equal to leading order.

In summary, $\overline{r}(p_\theta, p_z)$ and $\overline{v}_\parallel(p_\theta, p_z)$ are implicit functions determined from the minimum of the effective potential, which corresponds to the radial coordinate of the gyro center. It is not trivial in actual calculations to determine this minimum in terms of the separating invariants $(\varepsilon, p_\theta, p_z)$, compare Ref. [21]. A convenient property with the set $(\varepsilon, v_\parallel, \overline{r})$ is that it is not necessary to explicitly calculate the minimum of the effective potential, since the dependence of this set of invariants on the $(r, v_\parallel, v_\perp, \varphi_g)$ phase space variables are known beforehand, and any macroscopic quantity can be determined from velocity integrals by using the formula

$$\int_{-\infty}^{\infty} d^3 v = \int_{-\infty}^{\infty} dv_\parallel \int_0^{\infty} dv_\perp v_\perp \int d\varphi_g$$

## VII  Screw pinch Vlasov equilibria

The set of invariants $(\varepsilon, v_\parallel, \overline{r})$ is particularly flexible for a screw pinch to adjust the chosen distribution function $F(\varepsilon, v_\parallel, \overline{r})$ to expected or measured profiles of the physical quantities. In screw pinches, no mirror trapping appears, and $\mu = \mu(\varepsilon, p_\theta, p_z) = \mu(\varepsilon, v_\parallel, \overline{r})$ is not an independent invariant in this geometry [this result for screw pinches was not anticipated in Ref. 3, where it was incorrectly suggested that $(\varepsilon, \mu, p_\theta, p_z)$ would be a set of four independent invariants for a screw pinch]. We illustrate the procedure to construct a Vlasov equilibrium with a local Maxwellian distribution function

$$F(\varepsilon, \text{v}, \bar{r}) = n_0(\bar{r})[\frac{M/2}{\pi k_B T_0(\bar{r})}]^{3/2} e^{-\varepsilon'/k_B T_0(\bar{r})} \quad (22a)$$

where a Galileitransformed energy is introduced,

$$\begin{aligned}\varepsilon'(\varepsilon, \text{v}, \bar{r}) &= \varepsilon - q\phi_0(\bar{r}) - M\text{v}\,\text{v}_0(\bar{r}) + \frac{M}{2}\text{v}_0^2(\bar{r}) \\ &= q[\phi(r) - \phi_0(\bar{r})] + \frac{M}{2}\text{v}_\perp^2 + \frac{M}{2}[\text{v} - \text{v}_0(\bar{r})]^2\end{aligned} \quad (22b)$$

and the functions $n_0, \phi_0, T_0$ and $\text{v}_0$ correspond to radial profiles for the density, electric potential, temperature and mean velocity along **B**. The local thermal velocity for the ions and electrons are defined by $k_B T_{0,i}(r) = M\text{v}_{th,i}^2(r)/2$. A small gyro radius expansion yields [2,3] with $\bar{r} = r - r_g(\mathbf{x}, \mathbf{v})$

$$F(\varepsilon, \text{v}, \bar{r}) = F_c(\varepsilon, \text{v}, r) - \frac{\partial F_c(\varepsilon, \text{v}, r)}{\partial r} r_g(\mathbf{x}, \mathbf{v})$$

For a plasma with electrons and single charged ions, this gives the particle density

$$n(r) = n_0(r) \exp\{-\frac{q[\phi(r) - \phi_0(r)]}{k_B T_0(r)}\} \to n_0(r) \quad (22c)$$

where the result $\phi(r) = \phi_0(r)$ follows from quasi neutrality if the same functions $n_0(r)$ and $\phi_0(r)$ are chosen for both the ion and the electron distribution functions. The plasma current is [2,3]

$$\mathbf{j} = \sum_i \int_{-\infty}^{\infty} d^3\text{v}\,(q\text{v}\,F_c\hat{\mathbf{B}} - \frac{M\text{v}_\perp^2}{2B}\frac{\partial F_c}{\partial r}\hat{\mathbf{r}} \times \hat{\mathbf{B}})$$

where the last term is the diamagnetic drift arising from the pressure gradients. The parallel current $j(r) = n_0(r)\sum_i q_i \text{v}_{0,i}(r)$ is a summation over the contributions from the electrons and ions, and $|v_{0,i}/v_{th,i}| \ll 1$ for typical plasma parameters. Amperes law gives for the considered distribution function,

$$B_z(r) = B_0 - \mu_0 \int_r^\infty dr \, [\frac{B_\theta}{B} j + \frac{B_z}{B^2} \frac{dP}{dr}] \qquad (23a)$$

$$B_\theta^2(r) = -\frac{1}{r^2} \int_0^r dr \, r^2 \frac{d}{dr} [B_z^2 + 2\mu_0 P(r)] \qquad (23b)$$

where $P = n_0(r) \sum_i k_B T_{0,i}(r)$ is the total pressure. The last equation is identical to the radial force balance of ideal MHD.

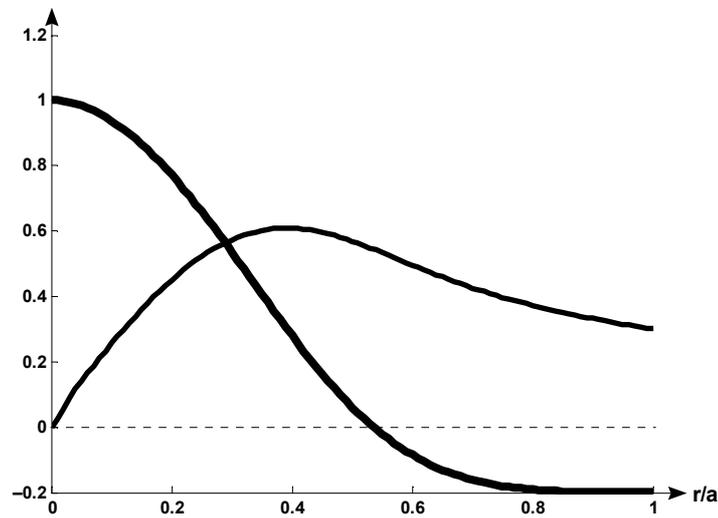

**Fig. 3** Representative reversed field pinch (RFP) profiles of normalized magnetic field components $B_\theta$ and $B_z$ with a reversal of $B_z$.

To illustrate how the distribution function can be used, a representative case is considered with the same temperatures for the ions and electrons and a zero mean parallel velocity for the ions, i.e. $v_{0,ion} = 0$. The RFP (reversed field pinch) model in Figs. 3 and 4 is computed with $\beta_0 = 0.2$, where $\beta_0 = 2\mu_0 P(0)/B^2(0)$. The electric field $E_r = -d\phi_0/dr$ is an arbitrary function in the model.

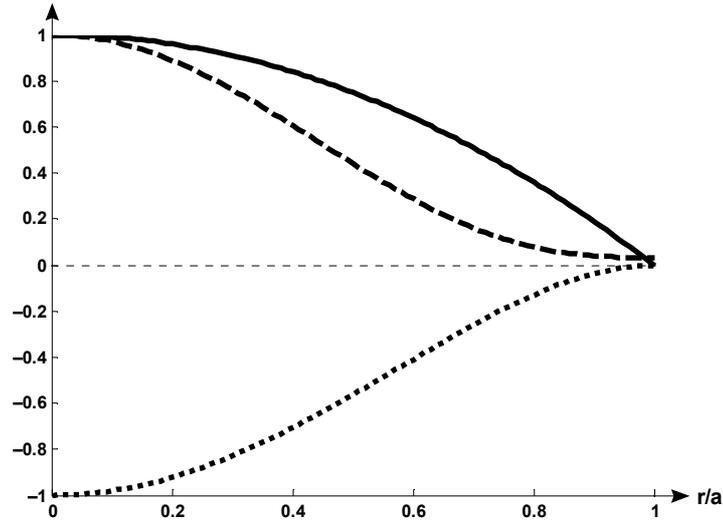

**Fig. 4**  Normalized density (solid line), temperature (dashed line) and mean parallel electron velocity (dotted line) for the same case as in Fig. 2.

## VIII  Linearized Vlasov equation for a screw pinch

The motion in a screw pinch is exactly periodic, and solutions of the linearized Vlasov equation can be found by constructing a periodic function of the gyro angle. Such solutions have been derived in a number of papers (see for instance Refs. [21,211 and 23]), so the approach in this section is not basically new. However, it could be interesting to see in some detail how the complexity of the analysis is reduced by using the set $(\varepsilon, v, \bar{r})$ of constants of motion and the harmonic oscillator approximation for the gyrating particle motion.

The perturbed quantities are Fourier decomposed as

$$f_1 = \int d\omega \int dk \sum_m e^{-i\omega t + im\theta + ikz} \hat{f}_1$$

The linearized Vlasov equation for the Fourier component becomes

$$-i(\omega - m\dot{\theta} - k\dot{z})\hat{f}_1 + v_r \frac{\partial \hat{f}_1}{\partial r} + \frac{dp_r}{dt} \frac{\partial \hat{f}_1}{\partial p_r} = \hat{h}$$

$$\hat{h} = -\frac{q}{M}(\hat{\mathbf{E}}_1 + \mathbf{v} \times \hat{\mathbf{B}}_1) \cdot \frac{\partial f}{\partial \mathbf{v}} \qquad (24)$$

Using the harmonic oscillator approximation for the unperturbed orbit, i.e. $r = \bar{r} + (v_\perp/\Omega)\cos\Omega\tau$, where $\Omega\tau = \varphi_g$ is the gyro angle, the small gyro radius expressions (20b,c) for $\dot{\theta}$ and $\dot{z}$ and the transformation $(r, p_r) \to (\tau, \varepsilon)$ yield

$$[-i(\omega - \omega_{m,k}) + i\Omega_{m,k}\frac{v_\perp \cos\Omega\tau}{\bar{r}\Omega} + \frac{\partial}{\partial\tau}]\hat{f}_1 = \hat{h}$$

where the drift terms and finite gyro radius terms are specified by

$$\omega_{m,k} = m\frac{\bar{v}_\theta}{\bar{r}} + k\bar{v}_z \qquad (25a)$$

$$\Omega_{m,k} = -m\Omega_\theta + k\bar{r}\,\Omega_z \qquad (25b)$$

Introducing the small parameter $\lambda = (\Omega_{m,k}v_\perp)/(\bar{r}\Omega^2)$ and

$$g(\tau) \equiv g(\varepsilon, \mu, v, \bar{r}, \tau) = (\omega - \omega_{m,k})\tau + \lambda \sin\Omega\tau$$

an integrating factor can be identified and the solution can be written

$$\hat{f}_1 = C e^{ig(\tau)} + \int_0^\tau d\tau'\, e^{ig(\tau)-ig(\tau')}\hat{h}(\tau') \qquad (26)$$

where $C$ is constant along unperturbed orbits. The solutions of interest are periodic in $\tau$ over a complete gyro period. The periodic solution for $-\pi \leq \varphi_g \leq \pi$ is obtained with

$$C = \frac{-e^{i\pi\alpha_{m,k}}\int_0^{\pi/\Omega}\hat{p}(\tau')\,d\tau' + e^{-i\pi\alpha_{m,k}}\int_{-\pi/\Omega}^0 \hat{p}(\tau')\,d\tau'}{2i\sin(\pi\alpha_{m,k})}$$

where $\alpha_{m,k} = (\omega - \omega_{m,k})/\Omega$ and $\hat{p}(\tau) = e^{-ig(\tau)}\hat{h}(\tau)$.

The periodicity can alternatively be described by a Fourier series in the gyro angle

$$\hat{f}_1 = \sum_p \hat{f}_{1,p} e^{ip\Omega\tau} \qquad (27a)$$

$$\hat{f}_{1,p} = -\frac{1}{i}\sum_{p',p''=-\infty}^{\infty}\frac{\hat{h}_{p'}J_{p''}(\lambda)J_{p-p'+p''}(\lambda)}{\omega-\omega_{m,k}-(p'-p'')\Omega} \tag{27b}$$

where $\hat{h} = \sum \hat{h}_p e^{ip\Omega\tau}$, $e^{i\lambda\sin\Omega\tau} = \sum J_p(\lambda)e^{ip\Omega\tau}$ and $J_p$ are ordinary Bessel functions of the first kind. In the zero gyro radius limit, this reduces to in the low frequency range

$$\hat{f}_{1,p} \to -\frac{1}{i}\frac{\hat{h}_p}{\omega-\omega_{m,k}-p\Omega} \approx -\frac{1}{i}\frac{\hat{h}_0 \delta_{p,0}}{\omega-\omega_{m,k}}$$

and

$$\hat{f}_1 \approx \frac{1}{i}\frac{1}{\omega-\omega_{m,k}}\frac{\Omega}{2\pi}\int d\tau \frac{q}{M}(\hat{\mathbf{E}}_1+\mathbf{v}\times\hat{\mathbf{B}}_1)\cdot\frac{\partial f}{\partial \mathbf{v}} \tag{28}$$

where the drift kinetic limit requires $|\omega-\omega_{m,k}| \ll |\Omega|$, and it can be noticed that $\omega_{m,k}$ is small for perturbations satisfying the resonance condition $mB_\theta(r)+k\bar{r}B_z(r)=0$ at a rational surface, since with a small perpendicular drift

$$\omega_{m,k} = \frac{m(B_\theta \overline{v} - B_z \overline{v}_\perp)}{\bar{r}B} + \frac{k\bar{r}(B_z \overline{v} + B_\theta \overline{v}_\perp)}{\bar{r}B}$$

$$\approx (mB_\theta + k\bar{r}B_z)\frac{\overline{v}}{\bar{r}B} - mB\frac{\overline{v}_\perp}{\bar{r}B_z}$$

and $-mB\overline{v}_\perp/(\bar{r}B_z) \approx -mE_r/(\bar{r}B_z)$ if the $\mathbf{E}\times\mathbf{B}$ drift is the dominating term for the drift.

## IX  Conclusions

The three established adiabatic invariants are separating invariants in the sense of Liouville. Additional independent adiabatic invariants have been shown to exist. The number of first order differential equations for the characteristics of the Vlasov equation determines the number of independent constants of motion. The number of independent constants of motion is reduced if a global symmetry associated with a cyclic coordinate is imposed on the system. Five independent constants of motion, including two generalized angular momenta, have been identified for a force free motion and the Hamilton Jacobi equation with nondegenerate frequencies.

For field free equilibria, the angular momenta provide a possibility to obtain nonuniform density profiles. If the distribution function would depend only on the separating invariants (i.e. the three velocity components), the density would be constant throughout the three-dimensional space.

A gyro helix invariant can exist for a gyrating particle, but that type of constant of motion is typically not relevant to use in a distribution function for fusion plasmas. We expect typically four "useful" invariants for a particle gyrating in a magnetic field. A confined gyrating particle has a constant mean value of the radial coordinate. A radial drift invariant can be identified from this property. To leading order the radial invariant is the guiding center radial coordinate, and the guiding center moves on a flux surface to this order. For the special case of a screw pinch, no magnetic mirror trapping appears and it turns out that the magnetic moment $\mu(\varepsilon, p_\theta, p_z)$ and the radial invariant are functions of the separating invariants. This is consistent with the result that the number of independent invariants is reduced by global symmetries. In a screw pinch, the set of constants of motion can be chosen either from the set $(\varepsilon, v, \bar{r})$ or from the separating invariants $(\varepsilon, p_\theta, p_z)$, but the former set is particularly convenient to model the equilibrium and to obtain a bridge between the Vlasov and fluid descriptions when the gyro radii are small. Local Maxwellian distribution functions expressed in this set of constants of motion have been demonstrated to provide MHD-type of equilibria. A harmonic oscillator approximation around the gyro center fluxsurface of the particle orbits yields a simplification of the linearized Vlasov equation, which could be useful to study resonance and orbit effects on stability.

A final remark could be made on the radial invariant as an adiabatic invariant. The analysis in this paper has been restricted to stationary fields. However, a confined particle would on average move on a magnetic flux surface also when slow time variations are present for the field components, provided the parallel electric field is small. This suggests that the radial drift invariant, i.e. the mean of the guiding center radial coordinate, should exist as an adiabatic invariant even when the fields vary slowly in time.

**ACKNOWLEDGMENTS**

The authors are grateful to Dr. Hans-Olof Åkerstedt at Luleå Technical University for discussions on finite gyro orbits. A discussion with Prof. Torbjörn Hellsten at the Royal Institute of Technology is acknowledged. Prof. Mats Leijon is acknowledged for support. The Swedish Institute has provided Vladimir Moiseenko with a grant to do studies at Uppsala University.